\documentclass[aps,nolongbibliography,superscriptaddress, twocolumn, showkeys, aps]{revtex4-2}
\usepackage[UKenglish]{babel}
\usepackage{latexsym}

\usepackage[pdftex, colorlinks=true, linkcolor=red, citecolor=blue, urlcolor=magenta, pdfstartview=FitH]{hyperref}
\usepackage{graphicx}	
\usepackage{amssymb}
\usepackage{amsmath}
\usepackage{xcolor}
\usepackage[normalem]{ulem}

\pdfoutput=1

\usepackage{bm}
\usepackage{verbatim}
\usepackage{amsmath}
\usepackage{amssymb}
\usepackage[utf8]{inputenc}

\usepackage{nicefrac}
\usepackage{braket}
\usepackage{pdfcomment}
\usepackage{pbox}
\usepackage{makecell}
\usepackage{hhline}
\usepackage{multirow}
\usepackage{tabularx}
\usepackage{soul}
\usepackage{float}
\usepackage{microtype}
\usepackage[dvipsnames]{xcolor}

\begin{document}
\title{Mismatch between Raman shear modes and ferroelectric polarization in 3R-MoS$_{2}$} 
%
%
\author{Johannes Schwandt-Krause}
\affiliation{Institute of Physics, University of Rostock, Rostock, Germany}
\author{Jan-Niklas Heidkamp}
\affiliation{Institute of Physics, University of Rostock, Rostock, Germany}
\author{Mohammed El Amine Miloudi}
\affiliation{Institute of Physics, University of Rostock, Rostock, Germany}
\author{Swarup Deb}
\affiliation{Institute of Physics, University of Rostock, Rostock, Germany}
\affiliation{Saha Institute of Nuclear Physics, Kolkata, India}
\affiliation{Homi Bhabha National Institute, Mumbai, India}
\author{Elisabeth Grune}
\affiliation{Institute of Physics, University of Rostock, Rostock, Germany}
\author{Kenji Watanabe}
\affiliation{Research Center for Electronic and Optical Materials, NIMS, 1-1 Namiki, Tsukuba 305-0044, Japan}
\author{Takashi Taniguchi}
\affiliation{Research Center for Materials Nanoarchitectonics, NIMS, 1-1 Namiki, Tsukuba 305-0044, Japan}
\author{Rico Schwartz}
\affiliation{Institute of Physics, University of Rostock, Rostock, Germany}
\author{Oliver K\"uhn}
\affiliation{Institute of Physics, University of Rostock, Rostock, Germany}
\author{Tobias Korn}
\affiliation{Institute of Physics, University of Rostock, Rostock, Germany}



\begin{abstract}
Sliding ferroelectricity in parallel-stacked two-dimensional van der Waals materials enables a broad range of novel device concepts, but exploiting it requires reliable, non-destructive assignment of the underlying stacking order and polarization state. Here, we combine Kelvin-probe force microscopy (KPFM) with low-frequency Raman spectroscopy to probe the polarization domains and stacking configurations of a exfoliated trilayer 3R-MoS$_{2}$ flake on a hBN substrate. We find that ABA and BAB — both stackings with zero net polarization — are indistinguishable in KPFM, yet show drastically different low-frequency shear modes. This observation is reproduced across multiple flakes and is corroborated by low-temperature photoluminescence. Notably, the standard bond-polarizability model does not account for the difference in shear-mode activity between the ABA and BAB configurations, indicating that the interlayer Raman response of these stackings is governed by physics beyond a simple polarizability picture. Our results show that none of the here-used individual techniques alone is sufficient to assign sliding-ferroelectric stacking order and motivate a combined spectroscopic–scanning-probe approach.

\end{abstract}

\maketitle

\section{Introduction}
The emergence of sliding ferroelectricity has added a genuinely novel route to switchable electric polarization. In van der Waals crystals built from individually non-polar layers, a vertical electric dipole can emerge purely from the way adjacent layers are stacked: when the interlayer registry breaks the out-of-plane inversion symmetry, charge is transferred across the interface and a spontaneous polarization appears that can be reversed by a small lateral slip between adjacent layers rather than by displacing ions within a layer.\cite{li2017binary,Li2021sliding,fox2023stacking} First demonstrated in parallel-stacked hexagonal boron nitride,\cite{yasuda2021stacking,Moshe2021sliding,woods2021charge} the same mechanism was soon established in rhombohedral (3R) transition-metal dichalcogenides (TMDCs),\cite{wang2022interfacial,rogee2022ferroelectricity} where the parallel stacking of the non-polar monolayers leaves a net out-of-plane polarization tied directly to the stacking sequence. 3R-MoS$_{2}$ has since become a model system for this physics, supporting electrically switchable and non-volatile polarization states,\cite{Ouyang2025ElectricallySwitching,Ye2024nonvolatile} cumulative and multi-state polarization in thicker stacks,\cite{Deb2022Cumulative,cao2024polarization} and spontaneous-polarization driven photovoltaic and piezoelectric responses,\cite{yang2022spontaneous,dong2023giant} making the reliable identification of its stacking order a prerequisite for both fundamental studies and device design.

Because the polarization is set by the local interlayer registry, a real flake is generally partitioned into stacking domains of opposite or differing polarization, separated by mobile domain walls whose motion underlies the switching itself.\cite{Ye2024nonvolatile,van2024engineering,sui2024atomic} Mapping these domains and assigning their stacking order is therefore central to the field, and a range of complementary probes has been developed for the purpose. Kelvin-probe force microscopy (KPFM) images the out-of-plane polarization directly through the associated surface-potential contrast,\cite{Deb2022Cumulative,yasuda2021stacking} while operando electron microscopy,\cite{ko2023operando} nano-photocurrent and plasmon imaging,\cite{zhang2023visualizing} and atomic resolution microscopy\cite{sui2024atomic} have resolved domain configurations and their dynamics. KPFM, however, is sensitive only to the \emph{net} out-of-plane polarization, so distinct stacking arrangements that carry the same polarization are expected to appear identical in such a measurement.

Optical spectroscopy offers a complementary, non-invasive window on stacking order~\cite{Jan2026}. Low-frequency interlayer Raman modes -- the rigid-layer shear and breathing vibrations -- are governed by the interlayer coupling and registry, and their frequencies and intensities serve as an established fingerprint of layer number, polytype, and stacking sequence in few-layer TMDCs.\cite{Puretzky2015LowFreq,Lee2016Polytypism,Na2018Davydov,vanBaren2019Stacking,Lu2015Rapid} The shear-mode activity in particular is well captured by the interlayer bond-polarizability model (BPM), which links the observed Raman response to the specific stacking configuration.\cite{Luo2015Stacking,Liang2017Interlayer} This sensitivity has recently been exploited to track stacking and ferroelectric switching in 3R-MoS$_{2}$ by shear-mode Raman spectroscopy,\cite{liang2023shear,ZilianYe2025Resolving3L,Liu2026ShearMode} and the polarization has likewise been shown to leave clear signatures in the excitonic and photoluminescence (PL) response of parallel-stacked MoS$_{2}$.\cite{ZiliangPRX,Deb2024,Moshe2025PL}

Most Raman-based polytype assignments rely on phenomenological or first-principles interpretation of multiple spectral features, with limited independent experimental validation. Here, we exploit the intrinsic ferroelectricity of the rhombohedral structure to cross-validate stacking assignments, combining low-frequency shear-mode Raman spectroscopy, KPFM, and low-temperature PL to study stacking domains in trilayer (3L) 3R-MoS$_{2}$. We find a counter-intuitive relationship between the ferroelectric and the vibrational pictures: domains that are indistinguishable in KPFM, because they share the same net ferroelectric polarization, are clearly separated by their shear-mode Raman fingerprints; a neutral domain shares its low-frequency vibrational signature with the polar domain rather than with its neutral counterpart. This contrast persists after thermal annealing and the rearrangement of the domain walls, and is reproduced across multiple flakes. Taken together, the complementary electrostatic, vibrational, and excitonic signatures uniquely resolve the stacking sequence, providing an experimentally anchored route to distinguish otherwise degenerate polytypes, with low-frequency shear-mode Raman proving sensitive to a structural distinction between 3R stacking domains that is not captured by a purely electrostatic probe such as KPFM.

\begin{figure*}
    \centering
    \includegraphics[width=\textwidth]{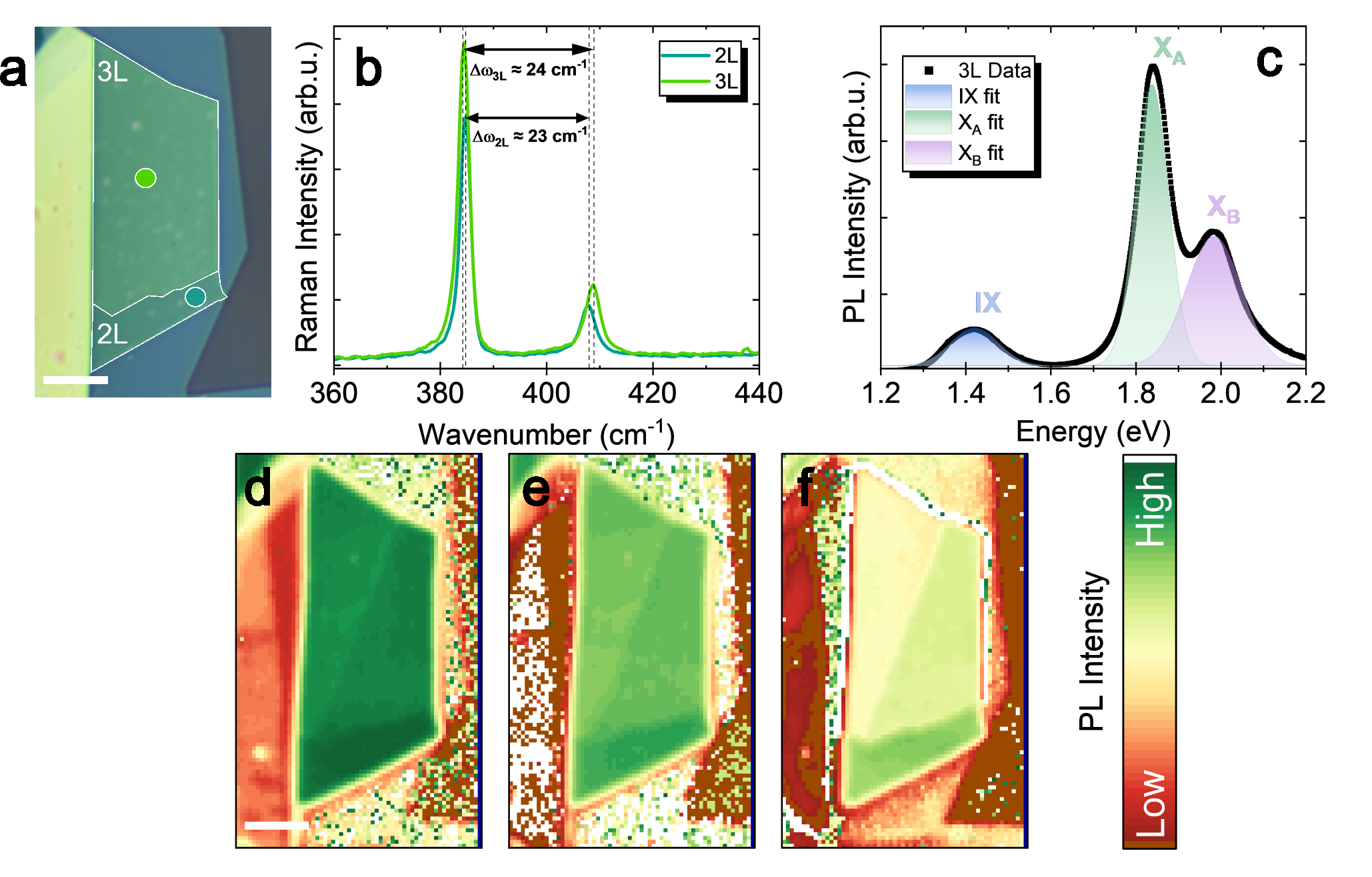}
    \caption{(a) Optical micrograph of the 3R-MoS$_{2}$/hBN flake showing the regions of differing layer number. The white bar denotes 5~µm. (b) High-frequency Raman spectra of the 2L and 3L regions, where the E$_{2g}$–A$_{1g}$ separation ($\Delta\omega_{\mathrm{2L}} \approx$ 23 cm$^{-1}$, $\Delta\omega_{\mathrm{3L}} \approx$ 24 cm$^{-1}$) confirms the layer assignment. The spectra are taken at representative coloured spots marked in (a). (c) Room-temperature photoluminescence spectrum deconvolved into the interlayer (IX), A-exciton (X$_{A}$), and B-exciton (X$_{B}$) contributions. (d-f) Spatially-resolved PL-intensity maps of the flake for the A-exciton (d), B-exciton (e) and interlayer exciton (f).}
    \label{Panel1}
\end{figure*}

\begin{figure*}
    \centering
    \includegraphics[width=\linewidth]{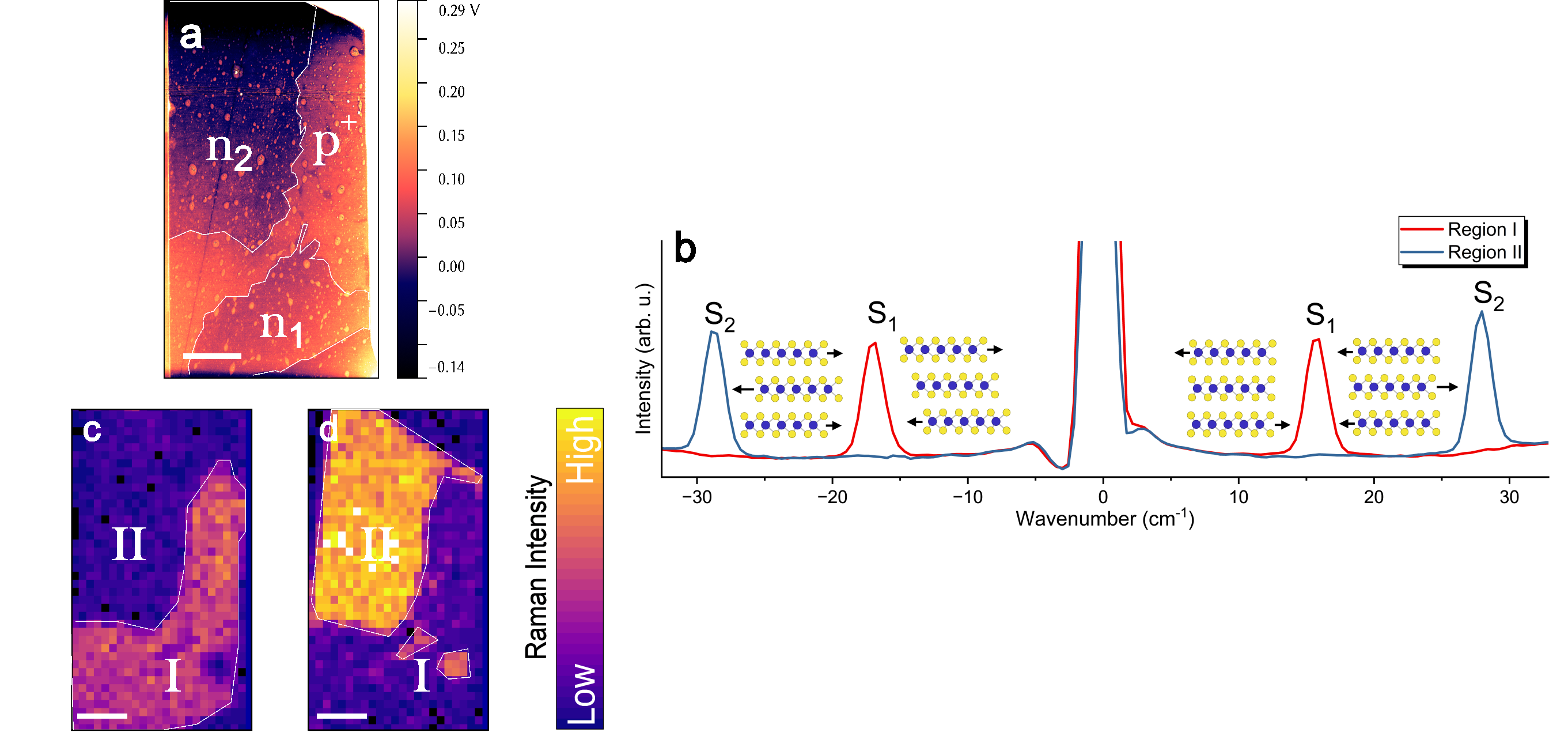}
    \caption{(a) KPFM surface-potential map of the 3L part of the MoS${_2}$ flake alongside Raman-intensity maps of the shear-modes. (b) Stokes/anti-Stokes low-frequency Raman spectra for two regions of the flake (Region I, red; Region II, blue), resolving the S$_{1}$ and S$_{2}$ interlayer shear modes. Inset crystal-structure schematics indicate the candidate stacking sequences and displacement associated with each region/shear mode. For the calculated structures, see Fig. S2, Suppl. Mat. (c) S$_{1}$ and (d) S$_{2}$, showing the spatial correlation between ferroelectric domains (n$_1$/n$_2$ denote neutral domains, p$^+$ denotes a polar domain) and shear-mode activity of different modes across the sample. The white bars denote 5~µm.}
    \label{Panel2}
\end{figure*}

\begin{figure*}
    \centering
    \includegraphics[width=\linewidth]{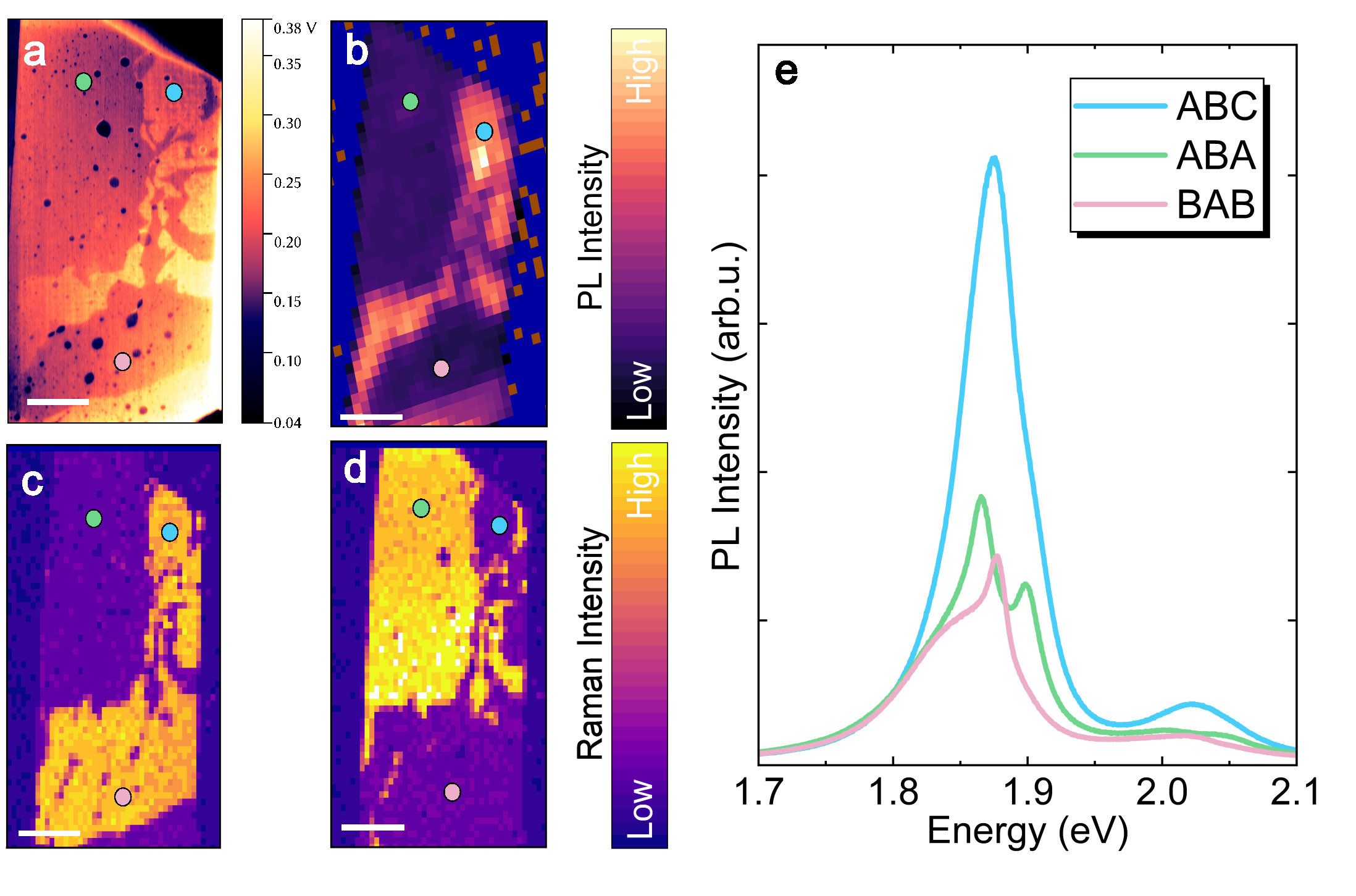}
    \caption{(a) KPFM surface-potential map and the corresponding low-frequency Raman-intensity maps of the shear modes (c) S$_{1}$ and (d) S$_{2}$  of a the same flake as in Fig. 1 and Fig. 2 after an annealing procedure. The white bars denote 5~µm. (b) Photoluminescence intensity map of the A exciton transition of the same area as in (c) and (d) at low temperatures (4~K). (e) Photoluminescence spectra  taken at three representative positions (ABC, ABA and BAB domain, respectively) show the variation in X$_{A}$ lineshape and the relative weight of the higher-energy feature between domains. The color of the plots matches the inset color markings shown in (a)-(d). Panel a is reproduced from ref~\cite{Jan2026}(published under a CC BY 4.0 license).}
    \label{Panel3}
\end{figure*}

\section{Results and discussion}
We begin by characterising a 3R-MoS$_{2}$ stack on top of a thin hBN layer utilizing high-frequency Raman and photoluminescence (PL) spectroscopy (see Supporting Information (SI), Sec. S2, for detailed information on fabrication and experimental methodology). The flake consists of adjacent 2L and 3L regions of 3R-MoS$_{2}$ (Fig.~\ref{Panel1}a) and is deposited on a thin ($\sim$10--20~nm) hBN layer on a 90~nm SiO$_{2}$/Si substrate. Figure~\ref{Panel1}b shows the in-plane E$_{2g}$ and out-of-plane A$_{1g}$ Raman modes for the 2L and 3L regions. Their separation increases with layer number, from $\Delta\omega_{\mathrm{2L}}\approx 23$~cm$^{-1}$ to $\Delta\omega_{\mathrm{3L}}\approx 24$~cm$^{-1}$, consistent with the well-established use of this mode spacing as a fingerprint for the layer count of few-layer TMDCs.\cite{Lee2016Polytypism,Na2018Davydov,LeeRamanLayerCount2010}

A representative room-temperature PL spectrum of the 3L region is shown in Fig.~\ref{Panel1}c. Three features can be resolved by their emission energy. The A and B excitons originate from the K/K$'$ points of the Brillouin zone and are intralayer in character, whereas the feature labelled ``IX'' at ~1.42~eV is commonly attributed to an interlayer~\cite{Deb2024,ZiliangPRX} or indirect~\cite{Zhao2013Indirect, Na2022Indirect} exciton, in which the electron is confined to a single layer while the hole is delocalised over several layers. The spatially resolved PL-intensity maps in Fig.~\ref{Panel1}d show that the emission intensity of all three features decreases with increasing layer count, which we attribute to the decrease of oscillator strength of the intralayer and interlayer excitons~\cite{splendiani2010emerging,Marie2016Lifetime}. Together, these measurements confirm the layer assignment and the overall quality of the flake.

Having established the layer assignment, we now turn to the stacking order, which we probe by combining Kelvin-probe force microscopy (KPFM), low-frequency Raman spectroscopy and low-temperature PL (Fig.~\ref{Panel2}). Figure~\ref{Panel2}a shows the KPFM surface-potential map of the 3L part of the flake. It distinguishes only two potential domains, divided in the surface-potential map into \emph{three} spatially distinct regions: a high-potential domain that dominates the right-hand side of the flake and extends through its centre, together with two low-potential domains: one occupying the upper-left to central region and one at the bottom -- that are offset from the high-potential domain by $\sim$110~mV. Each interface in a multilayer 3R stack contributes a polarization-induced potential step of approximately 60~mV, producing a characteristic potential ladder across the crystal.~\cite{Deb2022Cumulative,vizner2021interfacial,cao2024polarization} For the present setup this is verified by DFT calculations in the SI, Figs. S2 and S3. Accordingly, the four possible trilayer stackings — ABC, ABA, BAB, and CBA — yield three distinct surface potentials. ABC and CBA possess equal but opposite out-of-plane polarization, corresponding to the highest and lowest surface potentials, respectively, whereas the mirror-symmetric ABA and BAB stackings have zero net polarization and are therefore indistinguishable by KPFM. Consequently, the observed $\sim$110~mV potential contrast is consistent with two equally plausible assignments: either (i) the high-potential domain is ABC and the low-potential domains are ABA/BAB, or (ii) the high-potential domains are ABA/BAB and the low-potential domain is CBA. Thus, KPFM alone cannot uniquely determine the stacking sequence. We, therefore, turn to optical far-field probes that are sensitive to the stacking sequence itself rather than to its electrostatic fingerprint.

Low-frequency Raman spectroscopy provides exactly such a probe: the frequencies and Raman intensities of the interlayer shear modes are dictated by the stacking sequence, as established by early low-frequency Raman studies of few-layer TMDCs~\cite{Puretzky2015LowFreq,Lu2015Rapid,vanBaren2019Stacking} and described quantitatively by the interlayer BPM.~\cite{Luo2015Stacking,Liang2017Interlayer} Figure~\ref{Panel2}b shows two representative low-frequency spectra recorded on the 3L region in a cross-polarized configuration, in which the selection rules suppress the layer-breathing modes and leave only the interlayer shear modes visible. The two spectra correspond to distinct regions of the flake: region~I exhibits only the shear mode S$_{1}$, whereas region~II exhibits only the higher-energy shear mode S$_{2}$. Strikingly, the S$_{1}$ and S$_{2}$ intensity maps decisively lift the degeneracy inherent to the KPFM measurements by resolving the two spatially separated low-potential domains, which are otherwise indistinguishable in surface potential. Their distinct optical signatures establish that the two domains possess different stacking orders, leaving only one consistent assignment: the darker domains correspond to the net-neutral ($n$) ABA and BAB stackings, while the high-potential domain is uniquely identified as the polar ($p$) ABC stacking. Accordingly, we index the center brighter domain as $p^+$ and the two neutral domains as $n_1$ and $n_2$.

\par
Low-frequency Raman spectroscopy alone also fails to completely resolve the stacking domains (\ref{Panel2}c,d). While the S${_1}$ and S${_2}$ modes each provide stacking-sensitive contrast for ABA and BAB, it fails to distinguish the ABC domain from the net-neutral domain n${_1}$.
A key finding of this work is that the correlation between the KPFM potential contrast and the shear-mode Raman intensity is remarkably robust. Following thermal annealing, which is known to relax transfer-induced stress and redistribute ferroelectric domain boundaries, the domain landscape evolves substantially (Fig.~\ref{Panel3}a). Despite this rearrangement, the correspondence between the KPFM potential map and the Raman intensity remains unchanged. Most importantly, the shear-mode Raman maps (Fig.~\ref{Panel3}c,d) continue to distinguish the two net-neutral domains that are degenerate in KPFM, demonstrating that the Raman contrast is an intrinsic fingerprint of the local stacking order rather than a consequence of residual strain or other sample-specific inhomogeneities. The same KPFM–Raman correspondence is consistently observed in multiple independently prepared 3R-MoS${_2}$ flakes (Suppl. Mat., Fig. S8), establishing its reproducibility. Furthermore, the flake shown in Fig.~\ref{Panel1} was encapsulated after annealing to eliminate any influence of an asymmetric dielectric environment; the stacking-dependent Raman contrast remained unchanged (SI, Fig. S9). Together, these observations establish the shear-mode Raman response as a robust and intrinsic probe of stacking order in trilayer 3R-MoS${_2}$. Notably, this behaviour is at variance with previous studies, in which the shear-mode response of 3R-MoS${_2}$ was found to be fully consistent with the conventional BPM.~\cite{Liu2026ShearMode,liang2023shear,ZilianYe2025Resolving3L} In the SI, Section S1.E, we show results of  a DFT calculation of the Raman spectra, which are in very good agreement as far as the frequencies of the modes are concerned. DFT results also agree with the general predictions of BPM, thus confirming the local picture inherent to this approximate treatment. This means that DFT and BPM predict both ABA and BAB to have Raman activity in the S$_2$ mode only. This agreement is not surprising as the vanishing intensity of the S$_2$ mode is a consequence of symmetry selection rules. For further discussion, see Sec. S1.E in SI.

With stacking assignment in place, we now examine its excitonic signatures using low-temperature PL, which is highly sensitive to both the stacking sequence and the ferroelectricity induced band modulation in 3R-MoS${_2}$.\cite{Jan2026,Deb2024,Moshe2025PL,Schwause2026ILX} The spatial PL intensity map of the trilayer region Fig.~\ref{Panel3}b) closely follows the domain pattern observed by KPFM, indicating a strong correlation between the local electrostatic potential and the excitonic response. Figure~\ref{Panel3}e compares PL spectra acquired from three representative locations: the high-potential ABC domain (blue) and the two net-neutral domains, ABA (green) and BAB (pink). While ABA and BAB are electrostatically indistinguishable in KPFM, they exhibit distinctly different A-exciton spectral line shapes, demonstrating that the two mirror-related stackings possess markedly different excitonic responses.

To understand how two mirror-symmetric stackings of identical net polarization can nevertheless produce distinct excitonic spectra, we consider the microscopic mechanism identified for parallel-stacked MoS${_2}$ multilayers in Ref.~\cite{Deb2024}. Clearly, electrostatics alone cannot account for the observed contrast: the ferroelectric built-in potential shifts the layer-projected conduction and valence bands at the K points by the same amount, leaving the intralayer exciton transition energies unchanged. This is the case for ABC stacking leading to a narrower PL spectrum. The vertical potential sequence of the two neutral stackings is not a monotonic staircase-like structure. Instead it leads to resonant hybridization between the valence bands of electrostatically degenerate layers, namely the top and the bottom layers for ABA and BAB. The conduction-band states, on the other hand, largely retain their layer-localized character. The resulting unequal shifts of the layer-projected valence and conduction bands lift the degeneracy of the layer-hosted A-exciton transitions and imprint stacking-specific multiplicity or sub-features on the optical spectra. The strength of the K-point hybridization is set by the electrostatic potential of the middle layer, therefore, the interlayer registry of each interface. For ABA (BAB) stacking, where the central layer has a lower (higher) potential, the hybridization between the top and bottom layer is strong (weak). On this basis we tentatively assign the green spectrum, which resolves two dominant peaks separated by $\approx$34 meV, to the ABA stacking, in which the hybridization-induced splitting between the middle- and outer-layer-hosted A-excitons is resolved, and the pink spectrum, which shows a broad band underlying a single sharp peak, to BAB, where the corresponding splitting remains below the linewidth. We note that the asymmetric dielectric environment of vacuum/3L-MoS${_2}$/hBN geometry lifts the equivalence of the two outer layers and may contribute additional layer-dependent shifts and broadening, as discussed for symmetrically encapsulated and asymmetric trilayers in~\cite{ZilianYe2025Resolving3L,Moshe2025PL}. In a nutshell, the distinct line shapes of the two KPFM-degenerate domains independently corroborate their different stacking orders, consistent with the shear-mode Raman contrast seen in Fig.\ref{Panel3}c,d.

\section{Conclusion}
In summary, we have fabricated 3L-3R-MoS${_2}$ flakes on thin hBN layers and characterized their stacking and polarization states using a combination of complementary techniques. Standard high-frequency Raman and room-temperature photoluminescence spectroscopy confirmed the layer number and overall sample quality, while Kelvin probe force microscopy (KPFM) and low-frequency Raman spectroscopy enabled us to resolve the ferroelectric polarization domains and assign the interlayer stacking configurations in detail. Crucially, we observe that two neutral domains of the same 3L flake, which are expected to be vibrationally equivalent and therefore exhibit identical shear-mode responses, display distinctly different low-frequency Raman spectra. This finding indicates that KPFM and low-frequency Raman spectroscopy probe complementary, and not strictly redundant, aspects of the stacking and polarization landscape, and that the conventional assignment of stacking order in rhombohedral systems may be incomplete. Our results demonstrate that combining KPFM with low-frequency Raman spectroscopy yields a more complete picture of the stacking and polarization states than either technique alone, and motivate further systematic investigation of the stacking–polarization relationship in r-stacked two-dimensional materials.

\section{Acknowledgements}
T.K. acknowledges financial support by the DFG \emph{via} the following
grants: SFB1477 (project No.~441234705) and KO3612/8-1 (project No.~549364913).
K.W. and T.T. acknowledge support from the CREST (JPMJCR24A5), JST and World Premier International Research Center Initiative (WPI), MEXT, Japan.
The authors gratefully acknowledge technical assistance and fruitful discussions with J. Schr\"oer.

\bibliography{sn-bibliography}

\end{document}